\documentclass[manuscript]{aastex}

\slugcomment{}

\shortauthors {Stanghellini et al.}
\shorttitle {Planetary nebulae in DR2} 

\begin{document}

\title {The population of Galactic planetary nebulae: a study of distance scales and central stars based on the second GAIA release}

\author{Letizia Stanghellini}
\affil{NSF's National Optical-Infrared Astronomy Research Laboratory, Tucson, AZ 85719}
\email{lstanghellini@noao.edu}

\author{Beatrice Bucciarelli}
\affil{INAF, Osservatorio Astrofisico di Torino, Via Osservatorio, 20, 10025 Pino Torinese, Italy}
\email{}

\author{Mario G. Lattanzi}
\affil{INAF, Osservatorio Astrofisico di Torino, Via Osservatorio, 20, 10025 Pino Torinese, Italy}
\email{}

\and 
\author{Roberto Morbidelli}
\affil{INAF, Osservatorio Astrofisico di Torino, Via Osservatorio, 20, 10025 Pino Torinese, Italy}
\email{}

\keywords{Planetary nebulae: distances}

\begin{abstract}

We matched the astrometry of central stars (CSs) of spectroscopically-confirmed Galactic planetary nebulae (PNe) with DR2 Gaia parallaxes ($p$), finding 430 targets in common with $p>0$ and $|\sigma_{\rm p}/p|<1$. A catalog of PNe whose CSs have DR2 Gaia parallaxes is presented in Table~1.
We compared DR2 parallaxes with those in the literature, finding a good correlation between the two samples.  We used PNe parallaxes to calibrate the Galactic PN distance scale. Restricting the sample to objects with 20$\%$ parallax accuracy, we derive the distance scale ${\rm log}(R_{\rm pc})=-(0.226\pm0.0155)\times{\rm log}(S_{\rm H\beta})-(3.920\pm0.215)$, which represents a notable improvement with respect to previous ones. 
We found that the ionized mass vs. optical thickness distance scale for Galactic PNe is not as well constrained by the Gaia calibrators, but gives important insight on the nature of the PNe, and is essential to define the domain for our distance scale application.
We placed the CSs whose distance has been determined directly by parallax on the HR diagram, and found that their location on the post-AGB H-burning evolutionary tracks is typical for post-AGB stars.

\end{abstract}

\section{Introduction}
Planetary nebulae are essential probes of stellar evolution, being the gaseous and dusty remnant of asymptotic giant branch (AGB) evolution. When studied in populations, they are effective tracers of kinematic and chemical changes in the parent galaxy. Studies of PNe in the Milky Way have always been hampered by the difficulty of measuring their distances. Individually, only a handful of PNe has well-determined distances (see $\S$2). The rest of the Galactic PN distances have always been estimated through distance scales (e.g., Daub 1982; Cahn et al.~1992, hereafter CKS; Stanghellini et al.~2008, hereafter SSV; Frew et al.~2016, hereafter F16), where a few PNe with known distances have been used to constrain a physical relation between pairs of PN physical parameters, one of which to be distance-dependent.
We have studied distances to the CSs of 
Galactic PN using TGAS parallaxes after the first Gaia release (DR1), but the PN sample was very limited in DR1, and so was the distance scale study (Stanghellini et al.~2017, hereafter Paper I). 

The release of the second Gaia catalog (DR2) has prompted us to study the parallaxes $p$ of CSs of spectroscopically-confirmed PNe, and to review the most widely used PN distance scales using Gaia parallaxes as calibrators. Planetary nebula catalogs and Gaia parallaxes have been matched by other Authors since the DR2 release: The Gaia Collaboration (2018), in the context of showing HR diagrams with Gaia targets, has examined a selection of a few CSs of PNe on the stellar evolutionary diagram, filling the evolutionary gap between AGB and WD stars.  Kimeswenger \& Barr{\'{\i}}a (2018) have compared distances derived from DR2 to other distances and found reasonable agreement between the sets for short distances. Gonz{\'a}lez-Santamar{\'\i}a et al. (2019) have used Gaia parallaxes to estimate physical radii and expansion ages of PNe.

The mail goal of the present study is to use DR2 central star parallaxes to calibrate the Galactic PN distance scale.
 
In $\S$ 2 we summarize our search for Gaia matches to the CSs of Galactic PN. 
In this section we also compare Gaia DR2 parallaxes with other existing, reliable parallaxes and distances available to date, including some recent distances that had not been available at the time of previous distance-scale calibration, such as the expansion distances from {\it HST} images by Sc\"onberner et al.~(2018, hereafter SBJ). 

In Section 3 we calibrate the commonly used PN distance scales with the Gaia parallax measurements. In $\S$4 we present a limited study of the CSs and their location on the HR diagram, again using Gaia parallaxes. In $\S$4 we discuss the results of our study and future directions.

\section{Galactic PNe, their CSs, and DR2 parallaxes}

\subsection{The Galactic PN sample with reliable parallaxes from Gaia}

We initially matched the J2000.0 coordinates of spectroscopically-confirmed Galactic PNe (Acker et al.~1992; Kerber et al. 2003) with Gaia DR2 catalog
and found 655 unique Gaia targets that match the CSs with very high confidence. In this sample there are 497 stars with $|\sigma_{\rm p}/p|<1$, while the sample with $|\sigma_{\rm p}/p|<1$ and $p>0$ consists of 430 CSs.

Table~1 (published electronically in its entirety) includes all Galactic PNe whose CSs have a corresponding DR2 parallax. 
In column~(1) we give the PN name as in Simbad (the PN~G number), column~(2) gives the Gaia DR2 ID of the coordinate match; column~(3) gives the DR2 parallax and its uncertainty, in milli-arcseconds; 
column~(4) gives the nebular morphology, column~(5) the angular radius, in arcsec; columns~(6) and (7) give the logarithmic $H\beta$ flux and extinction constant and their uncertainties in cgs units; column~(8) gives the linear radius of the nebula in parsecs, calculated from the angular radius and parallax; Finally, column~(9) gives the references for the ancillary data, as explained in the table note.

Obviously, the sample of Galactic PNe with Gaia parallaxes represents an incomplete sample, which markedly declines beyond $\sim$2.5 kpc. This result should be taken into account when using this sample to derive population-broad results.

\subsection{Parallaxes of Galactic PNe from Gaia compared to other independent parallaxes and distances}

There are 25 PNe in DR2 whose parallaxes or distances had been previously measured with other reliable methods. In 
Fig.~1 (upper panel) we compare directly DR2 parallaxes with other trigonometric parallaxes (Harris et al. 2007, hereafter H07), and also with parallaxes derived by inverting other independent distances. Parallaxes used in Fig.~1 are given in Table~2. 
Distances by SBJ have been measured via multi-epoch expansion parallax and {\it HST} images, but suffer from indetermination of the expansion velocity of the PN shells. Distances via spectroscopic parallaxes, such as those of Ciardullo et al. (1999, hereafter C99), suffer from model dependency. 

The correlation between the DR2 and independent parallaxes is tight, with linear correlation index $R_{\rm xy}=0.98$. In the lower panel of Fig.~2 we show the residuals, whose average is 0.22 for the whole sample. 
This comparison represents a positive assessment of Gaia parallaxes for CS PNe, making them ideal probes to calibrate the distance scale. 
  \begin{figure}
   \centering
  \includegraphics[width=12truecm]{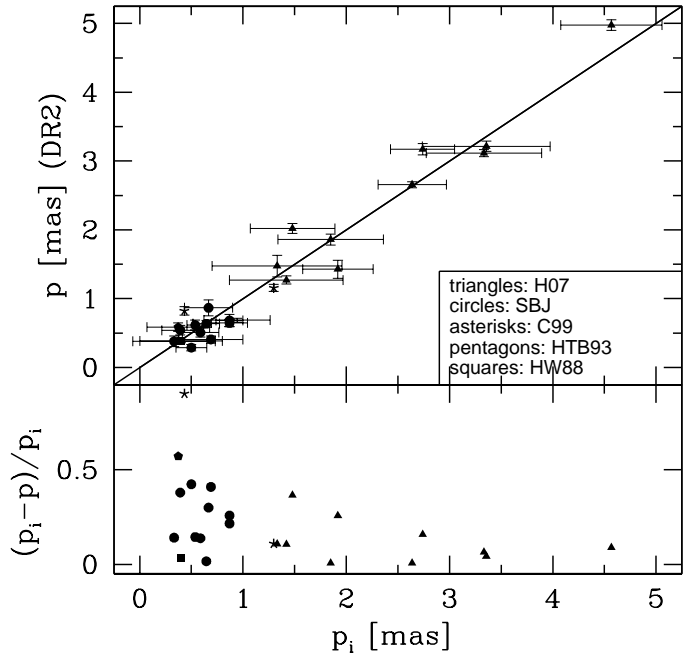}
\caption{Upper panel: Comparison between Gaia parallaxes and other independent parallaxes (or distances). References are H07: Harris et al. (2007); SBJ: Sc\"onberner et al. (2018);  C99: Ciardullo et al. (1999); HTB93: Hajian et al. (1993); HW88: Huemer \& Weinberger (1988). The solid line is the 1:1 relation. The lower panel shows the residuals. }
    \end{figure}

 \section{Galactic PN distance scale calibrated with Gaia parallaxes}

\begin{figure}
   \centering
  \includegraphics[width=12truecm]{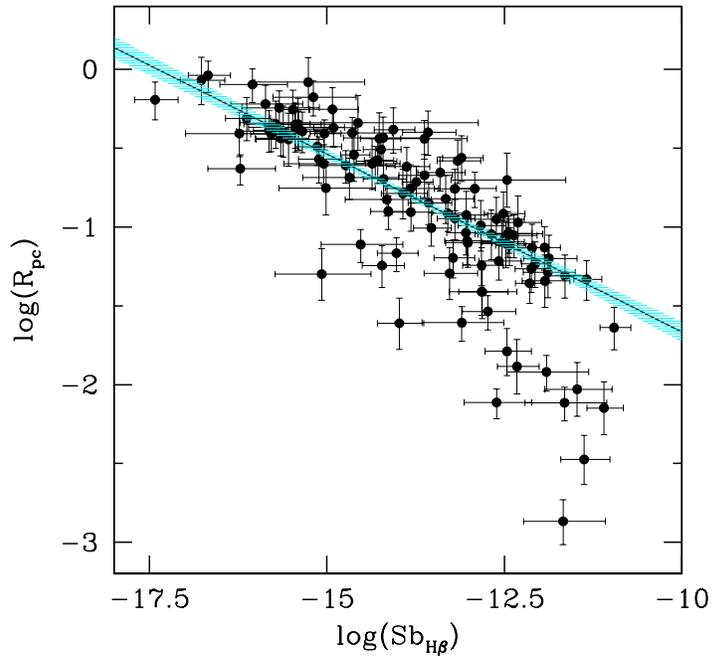}
\caption{The physical radius vs. ($H\beta$) surface brightness relation, as defined by PNe with DR2 parallax uncertainties smaller than $20\%$ (plotted with their uncertainties as solid circles). The solid line is the maximum likelihood relation of Eq.~(1). The shaded (light blue) area represents the $1\sigma$ confidence interval.}

\end{figure}

We set to use the DR2 parallaxes to study and calibrate the statistical distance scales for Galactic PNe. Statistical distance scales for Galactic PNe are based on the physical relation between distance-dependent and distance-independent parameters that are commonly measured or derived in PN studies. Such relations, once calibrated on PNe with known distances, are used to infer the statistical distances to all other PNe, whose distance is not known. As seen in the previous section, before the release of Gaia DR2 there were only 25 PNe whose distances were reasonably well known, half of whose became available only recently, thus the calibrator sample was really limited.
The DR1 Gaia release included a handful of calibrators, which we examined in Paper I, with the conclusion that the physical radius vs. surface brightness statistical distance scale was promising. The DR2 Gaia release gave us high motivation to re-explore PN distance scales. The next section includes a study on the re-calibration of the Galactic PN distance scales based on DR2 parallaxes. As in Paper I, we studied the distance scales relating (i) the physical radius to surface brightness and (ii) the ionized mass to optical thickness.

\subsection{The physical radius vs. surface brightness distance scale}

 This scale is realized by the assumed linear relation between ${\rm log}(R_{\rm pc})$ and ${\rm log}(Sb_{\rm H\beta})$, representing the distance-dependent and distance-independent parameters, respectively. The physical radius in parsec, $R_{\rm pc}$, is computed as $\theta/(206265\times p)$, where $\theta$ is the angular radius and $p$ the DR2 parallax, both in arcsec; the $H\beta$ surface brightness is computed as ${\rm log}(Sb_{\rm H\beta})= {\rm log} (I_{\rm H\beta}/\pi \theta^2$), where ${\rm log}(I_{\rm H\beta})={\rm log}(F_{\rm H\beta})+c$ is the extinction-corrected logarithmic $H{\beta}$ flux. 

The sample of Fig.~2, consisting of 112 targets,  shows the distribution in this plane of PNe whose parallax is available from DR2 with accuracy better than $20\%$, and whose physical parameters used for the distance scale are also available (see Table~1). The error bars in Fig.~2 reflect the propagation of uncertainties of all parameters. With this pruning, we retain a quite sizable distribution of calibrators whose parallax uncertainty compare favorably with the ones expected from the cosmic scatter of the statistical scale (see Buckley \& Schneider 1995; Schneider \& Buckley 1996), making them physically more revealing. 
 
By limiting $|\sigma_{\rm p}/p|$ we could, in principle, introduce a Lutz-Kelker-type bias which would principally affect the upper-left end of the 
${\rm log}(R_{\rm pc})- {\rm log}(Sb_{\rm H\beta})$ relation, where
 larger distances play a role. However, in the present analysis we did not take it into consideration, given that our calibration is also affected by
the intrinsic scatter of the ${\rm log}(R_{\rm pc})- {\rm log}(Sb_{\rm H\beta})$ relation, and by the presence of some stragglers, probably evolutionary in nature, populating the lower-right end of the ${\rm log}(R_{\rm pc})- {\rm log}(Sb_{\rm H\beta})$ plane. All these effects could be better investigated using future Gaia data releases. 

A simple weighted least squares fit in presence of asymmetric error bars would not suffice for this distance scale. The issue here is that we measure the parallax (of the CSs), the PN radius, and the PN line flux and extinction for the calibrators, and uncertainties, but we do not measure directly distances, physical radii, and surface brightness. Thus we approached the problem by utilizing only the measured quantities and their original uncertainties, to calibrate the distance scale. An optimal way to perform this calibration is a maximum likelihood approach, which is illustrated in Appendix A, and used to estimate the parameters of the PN statistical scale (see Kelly 2007). 

In Fig.~2 we show the maximum likelihood result as a thick solid line, corresponding to the following fit:
$${\rm log}(R_{\rm pc})=-(0.226\pm0.0155)\times{\rm log}(Sb_{\rm H\beta})-(3.920\pm0.215) \eqno(1).$$ 

 The distribution of PNe around this linear scale presents considerable scatter. This parameter distribution -- not observed in the TGAS sample (Paper I) due to the paucity of objects-- besides depending to some extent on the scatter in the measured PN parameters, is an empirical evidence that not all evolving PNe follow a unique physical correlation on this plane. 
 
To gain insight on the actual line (or strip) that PNe follow in their evolution, we needed to analyze parameter dependency.
In Fig.~3 we show the same data distribution of Fig.~2, but we set in evidence specific groups of PNe. We differentiated PNe with small apparent radii ($\theta<3\arcsec$, upper left), low radiation intensity (${\rm log}(F_{\rm H\beta})<-13$, upper right), high extinction constant ($c>1.5$, lower left), and low ionized mass (${\rm log}(\mu)<-2$, lower right, see next section for definition).  These are the critical PN group for which the errors may be higher than reported in the literature. We found that, by eliminating these extreme groups of PNe, the correlation of the distance scale is tighter. The parameter limits selected for Fig.~3 are quite arbitrary, and were used to recognize that the PN distances derived from the physical radius vs. surface brightness relation (Eq.~1) may have larger uncertainty for PNe observed very early in their evolution (smaller radii), or large extinction, or low $H\beta$ fluxes. The distribution indicates that the relation is poorly followed by PNe with very low ionized masses, as the ones which are still in an ionization-bound state. 

It is worth noting that the angular radii available in the literature have been measured in different ways. The best way to measure angular radii is by photometry, defining the photometric radii as the one encompassing 85$\%$ of the  total nebular flux. This method has been used almost exclusively when {\it HST} images of the PNe where available. To date, the only homogeneous set of photometric radii of Galactic PNe published so far being the one by Stanghellini et al. (2016), which is limited to small PNe, unresolved from the ground. Given the importance of a larger data set with photometric measured data, we measured the PN sizes of an additional sample of PNe whose {\it HST} images are available on the data archive.   We have performed aperture photometry on these images at various distances from the center, which corresponds to the location of the central star or to the geometrical center of the PN if a central star is not seen in the image. We then measure the total flux of the PN with broad aperture photometry, and find a radius that encircles 85$\%$ of total flux. The technique used here is identical to that of Stanghellini et al. (2016). The resulting radii are given in Table~1 with reference code {\it 1}.

In Fig.~4 we plot the physical radius vs. ($H\beta$) surface brightness locus for the $|\sigma_{\rm p}/p|<0.2$ sample,  i.e., the same data set of Fig.~2, where we have put in evidence the sample of extended PNe with photometric radii measured from {\it HST} images, who define a tight correlation on this plane ($R_{\rm xy}=-0.92$).  The PN sample with photometrically measured radii is too small to derive accurate coefficients through maximum likelihood analysis, and thus we will not give its fit, but we can see from Fig.~4 that these 8 targets define a much tighter correlation than the general sample, showing the importance of measuring photometric radii. Note that these PNe have different morphologies. With the final Gaia release, and with future {\it HST} observations, there may be more of these calibrators available for the distance scale. 

We explored the distance scale with calibrator PNe of different morphological types.  From the data set of Fig.~2 we have selected PNe that have been morphologically classified, and plotted them in Fig.~5 on the same plane than in previous figures, omitting error bars for clarity.  In this paper we used the classification scheme by Stanghellini et al. (1993) with {\it Round}, {\it Elliptical}, {\it Bipolar Core}, {\it Bipolar}, and {\it Pointsymmetric} PNe. Morphological classification is  given in Table~1, based on images published by Manchado et al. (1996), Schwarz et al. (1992), and  Balick (2007).  The sample of Fig.~5 includes 16 {\it Round} PNe,  with linear correlation coefficient  $R_{\rm xy}=-0.97$ in the scale plane. This was expected from the result of Fig.~4, since the photometric and geometric radii do correspond in {\it  Round} PNe. 

With the maximum likelihood method we determine the fit to {\it Round} PNe whose $|\sigma_{\rm p}/p|<0.2$ to be:

 $${\rm log}(R_{\rm pc})=-(0.267\pm0.0365)\times{\rm log}(Sb_{\rm H\beta})-(4.45\pm0.498). \eqno(2)$$ 

If we extend the sample to {\it Round} PNe with $|\sigma_{\rm p}/p|<1$ we found a similar correlation, ${\rm log}(R_{\rm pc})=-(0.253\pm0.0365)\times{\rm log}(Sb_{\rm H\beta})-(4.24\pm0.490)$, which is compatible with the fit of Eq.~(2) within the errors. 

Our study assumes that $\sigma_{\theta}/\theta=0.2$. 
 Errors on radii measurements are generally not published, and we used this  conservative estimate in our likelihood analysis of the scale parameters.
Biases in these measurements can arise if the PN is not spherically symmetric. In fact, our Fig.~5 shows that {\it Round} PNe are much less dispersed about the scale relation than the rest of them.
It would be difficult, with the sample at hand, to quantify the influence of such possible biases; the differences between the fits of Eqs.~(1) and (2) could be attributed partly to such an effect, though they are still compatible with the stated statistical uncertainties on theta.

We evaluate the accuracy of the scales represented by Eqs.~(1) and (2) with the method presented by Smith (2015, hereafter S15). We calculate the distance ratios $K$\footnote{Note that S15 calls distance ratio $R$, but we want to avoid confusion with the physical radius used in our scales.} by multiplying distances from the distance scales  -- excluding in all cases PNe with ${\rm log}(\mu)<-2$, which do not follow our physical scale -- to DR2 parallaxes. We used Eq.~(1a) in S15 to estimate the variance of the distance ratios, $\sigma_{\rm K}^2$, and Eq.~(5) for the typical relative errors, $\alpha_{\rm s}$. 
We used the scale distances $d_{\rm s}$ and DR2 parallaxes $p\pm\sigma_{\rm p}$ directly in Eq.~(1a), while we estimated $\sigma_{\rm s}^2$ for the scale distances by propagating the formal errors of Eqs.~(1) and (2) and accounting for correlations also in the observed parameters.

In Table~3 we give the results of our accuracy evaluation for Eqs.~(1) and (2), both compared to DR2 parallaxes, for ${\rm log}(\mu)>-2$. We also calculate these statistics for {\it Round} PNe exclusively, for both scales, to determine the best one to use in each case. In the first three columns of Table~3 we give the scale used for distances, the sample studied, and the sample population. Column~(4) gives the typical error of the scale, $\alpha_{\rm s}$; columns~(5) and (6) the average distance ratio $<K>=<d_{\rm s}\times p>$, and the average variance of the distance ratio, $<\sigma_{\rm K}>$.  In this analysis, the best distance scale is with K as close as possible to unity, and with the smallest $\sigma_{\rm K}$, since $\sigma_{\rm K}$ represents the fractional uncertainty in the distances from the studied scale.

If we compare the scales with parallaxes without considering the nebular morphology, we find that Eq.~(1) gives a slightly short scale ($-2\%$) with respect to Gaia parallaxes, while Eq.~(2) gives a slightly long scale ($+7\%$). It makes more sense to asses the scales with the best parallaxes, i.e., $|\sigma_p/p|<0.2$; in these cases, $<\sigma_{\rm K}>\sim0.25$. By limiting the comparison to {\it Round} PNe both distance scales are very accurate. 

The best results in Table~3 are those obtained from comparing Eq.~(1) and Eq.~(2) with the sample used to derive them, i.e., Eq.~(1) with all objects with $|\sigma_{\rm p}/p|<0.2$ and Eq.~(2) with {\it Round} objects with $|\sigma_{\rm p}/p|<0.2$. Also, it is worth noting that Eq.~(1) works as well for the sample used to define it as for {\it Round} PNe with secure parallaxes. Finally, in principle, the mean of Eq.~(1) and Eq.~(2) applied to {\it Round} PNe with secure parallaxes will be almost entirely free of bias. 

In terms of average distance ratios, i.e., by comparing the distances from a statistical scale with DR2 parallaxes, both Eqs.~(1) and (2) are a better scale than those published previously. For example, $<K>=1.942\pm2.671$ if calculated with distances from F16 and DR2 parallaxes with $|\sigma_{\rm p}/p|<0.2$,  based on the 128 PNe available for this comparison.  By limiting the comparison between F16 distances and DR2 parallaxes to PNe with ${\rm log}(\mu)>-2$ we obtain $<K>=1.217\pm0.513$ (95 PNe with $|\sigma_{\rm p}/p|<0.2$  available for this comparison), which is still indicative of a long scale.

We also performed a direct comparison between our scales and other widely used distance scales by using the $k=\Sigma_{\rm i} d_{\rm y, i}/\Sigma_{\rm i} d_{\rm x, i}$ estimator by Phillips (2002, see also in Eq.~(2) in S15; note that $k$ is different from the above $K$), where $d_{\rm y, i}$ is the distances of the i-th object in the old distance scale, and  $d_{\rm x, i}$ is the distance of the same object in our scale. By comparing F16 distances with our scales we found $k=1.243$ and $1.165$ respectively for Eqs.~(1) and (2). Both comparisons are based on the 103 PNe in common. The same comparisons with the SSV distances yielded $k=1.136$ and $k=1.068$. 

Our scale of Eq.~(1) should be used when DR2 Gaia parallaxes are not known. 

\begin{figure}
   \centering
  \includegraphics[width=12truecm]{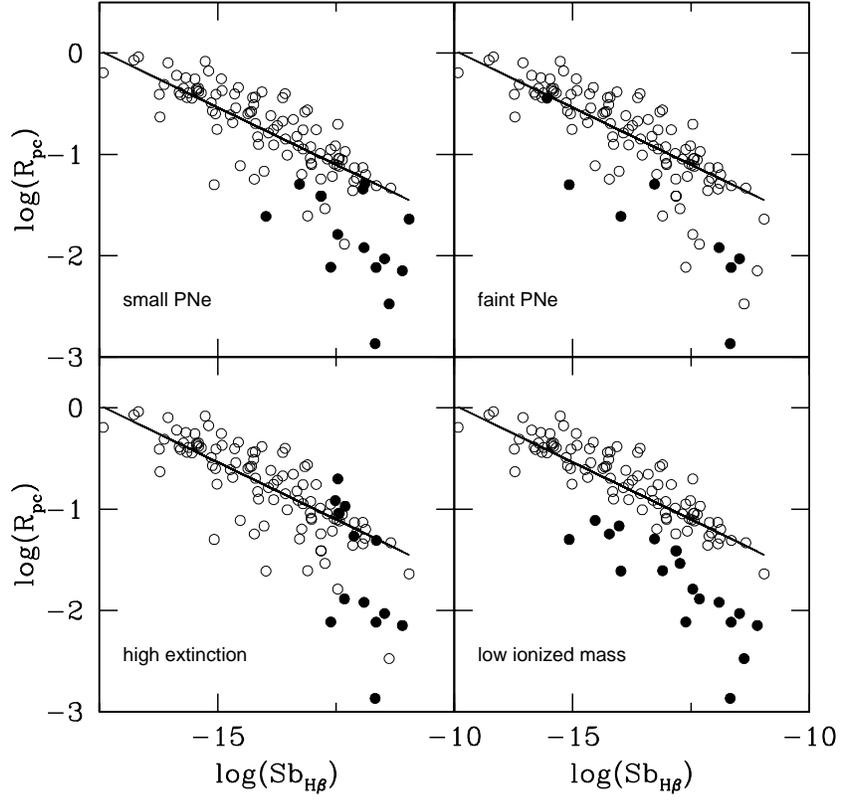}
\caption{Upper left: $\theta<3\arcsec$; Upper right: $F_{\rm H\beta}<10^{-13}$, Lower left:  c$>$1.5. Lower right: ${\rm log}(\mu)<-2$. The solid lines are the fit in Fig.~2, and Eq.~(1). error bars have been eliminated for clarity.}
\end{figure}

\begin{figure}
   \centering
  \includegraphics[width=12truecm]{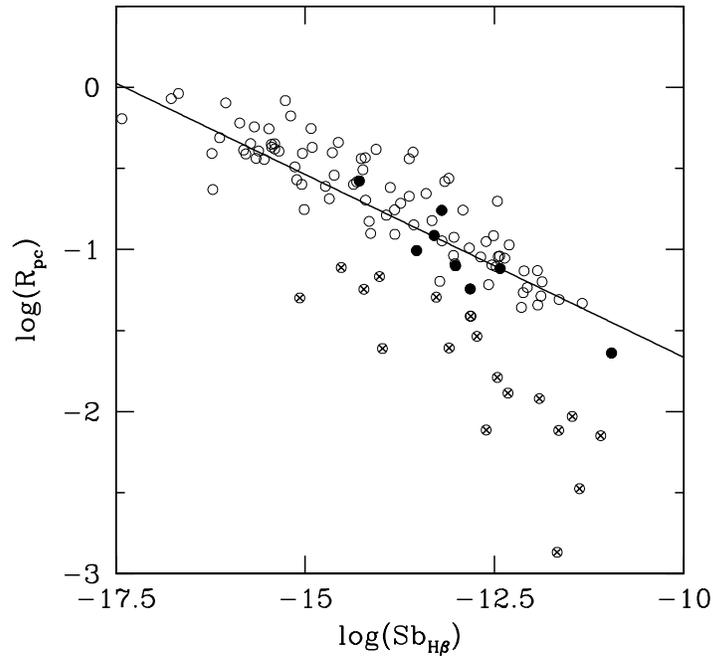}
\caption{The physical radius vs. ($H\beta$) surface brightness relation of Fig.~2, (i.e., with $|\sigma_{\rm p}/p|<0.2$) where indicate with filled symbols the extended PNe with photometric radii from this study (see Table~1), and with crossed symbols the PNe with ${\rm log}(\mu)<-2$. The straight line represents Eq.~(1).}
\end{figure}

\begin{figure}
   
 \centering
  \includegraphics[width=12truecm]{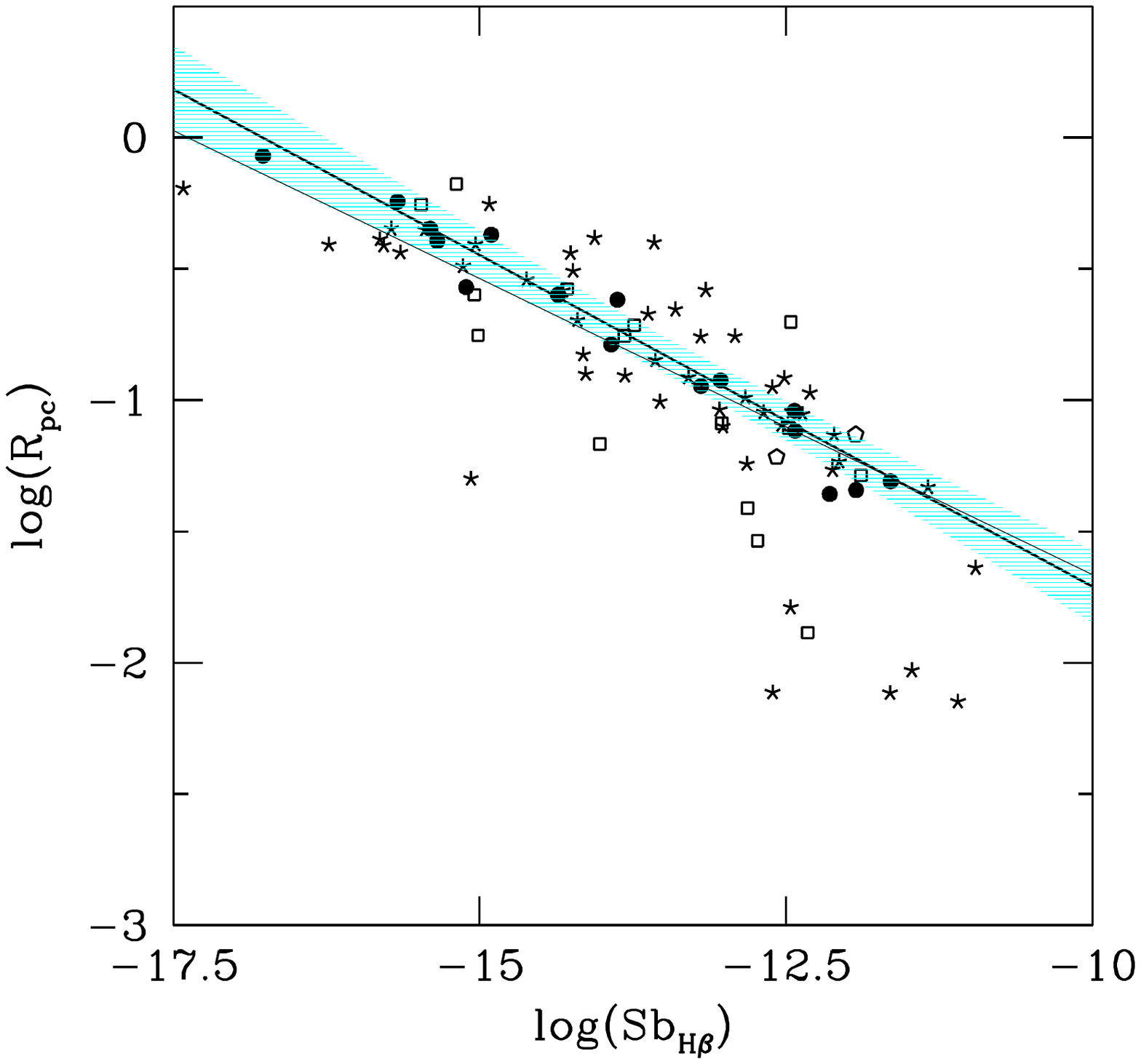}
\caption{The physical radius vs. ($H\beta$) surface brightness relation for the subset of Fig.~2, plotting only those PNe with known morphology. Asterisks: {\it Elliptical}; circles: {\it  Round}; squares: {\it Bipolar or Bipolar Core}; pentagon: {\it Pointsymmetric}. The solid thick line is the correlation found for  {\it Round} PNe with $|\sigma_{\rm p}/p|<0.2$, while the shaded area has the same meaning as in Fig.~2. The dark straight line is Eq.~(2); the light straight line is Eq.~(1), drawn for comparison.}
\end{figure}

\subsection{The ionized mass vs. optical thickness distance scale}

In Fig.~6 we show the relation between the ionized mass and optical thickness with PNe with low parallax uncertainties (i.e., $|\sigma_{\rm p}/p|<0.2$). A similar plot has provided the basis for the distance scales of Galactic PNe by CKS, and by SSV, where the former has been calibrated with Galactic PNe, and the latter with Magellanic Cloud PNe. Note that the domain of calibrators in those published scales was limited to the parameter space defined by the small rectangle in the Figure. The concept of this scale is that PNe evolve from optically thick to optically thin, and once reaching the optically thin (or density bounded) status, their ionized mass is constant. This scale includes the strong assumption that all PNe have identical ionized mass, and that they evolve in the same way from optically thick to optically thin. 

The abscissa, $\tau={\rm log}(4\times\theta^2/F)$, is a measure of the inverse optical thickness of the nebula, where $\theta$ is the angular radius and $F$ the 5 GHz flux of the PN. The ordinate, $\mu=(2.266\times10^{-21} p^{-5}\theta^3F)^{0.5}$, the ionized mass, is the distance-dependent term that has been calculated with $p$ (the Gaia parallax in arcsec) and the 5 GHz fluxes for the data points of Fig.~6 (5 GHz fluxes from CKS; Acker et al. 1992). 

 The optically thick sequence in the figure is represented by the sloping segment of the solid and broken lines, while the optically thin sequences are the horizontal lines. Neither sequence is well defined by the Gaia calibrators (i.e., when using $p$ in the calculation of the ionized mass). 
 
The optically thin sequence is reasonably reproduced by {\it Round} PNe, identified in the Figure as filled symbols. The average dispersion of ${\rm log}(\mu)$ are -0.931$\pm$0.72 and -0.81$\pm$0.39 respectively for the whole sample and the  {\it Round} PN sample. This finding agrees with the spherical assumption of the ionized mass term in CKS's calibration  and shows that this scale may not be accurate for other PN morphologies. The data set is too small to make a meaningful calibration of this distance scale based on photometric radii.
 
We note that PNe with ${\rm log}(\mu)<-2$ seem to lie in a parallel sequence where the ionized mass is underestimated, or they simply evolve differently from the others. If we plot the PNe with  
 ${\rm log}(\mu)<-2$ on the ${\rm log} (R_{\rm pc})$ vs. ${\rm log}(Sb_{\rm H\beta})$ scale we find that they are all located far from Eq.~(1) (see Fig.~3, lower right panel).
 
PNe with such low ionized mass are a minority, representing $\sim20\%$ of the whole population. We explored the parameter space for these PNe. We showed in Fig.~3 that several of them have a combination of small apparent radii, low fluxes, and high extinction although not all high extinction PNe have low ionized mass. Filling factors for these PNe can not be calculated, due to the lack of electron density and temperature information in the literature. The distance scale of Eq.~(1) calculated exclusively for PNe whose ${\rm log}(\mu)>-2$ is within the 1$\sigma$ of Eq.~(1) (i.e., within the shaded area of 
Fig.~2).
 
Since optical thickness varies with metallicity (see SSV) we also inspect their metallicity relative to solar through their oxygen abundances (from Stanghellini \& Haywood 2018), and found that most of these targets are oxygen-rich (supersolar). Given that the low-ionized mass PNe are not exclusively supersolar, since several targets with ${\rm log}(\mu)>-2$ are also supersolar, we could not conclude this to be the only factor for straying from the distance-scale correlations.  Additional analysis of these targets is in order for more insight. 
 
Finally, we noted that low-extinction PNe populate exclusively the optically thin sequence. Note that a similar result was found by  Kimeswenger \& Barria (2018); in fact, by comparing distances from Gaia DR2 parallaxes and those by SSV they found that reddened objects compare worse than blue objects for targets with distance $<$4 kpc. 

\begin{figure}
   \centering
  \includegraphics[width=12truecm]{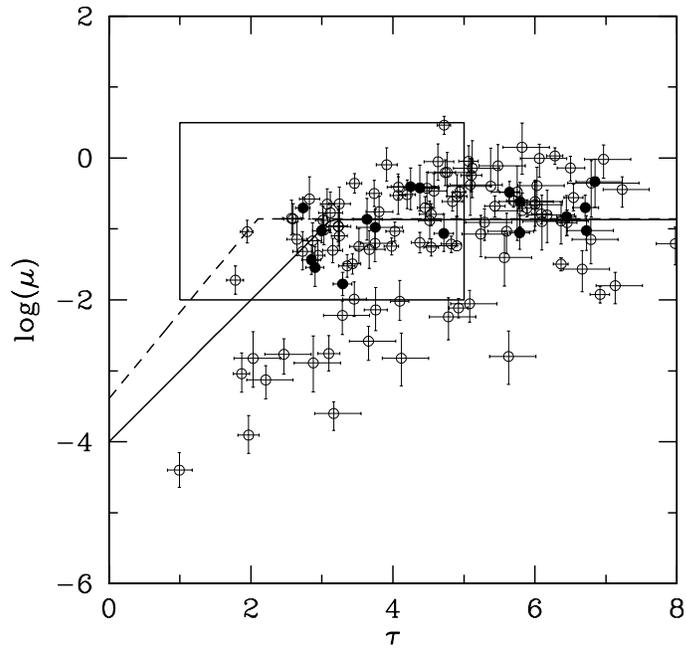}
\caption{The ionized mass vs. optical thickness relation.  The solid line is the CKS distance scale, while the broken line is the SSV scale, calibrated on Magellanic Cloud PNe. The box represents the extent to which the relation was populated when the CKS and SSV scales where derived. The solid circles are the PNe with  {\it Round} morphology.}
\end{figure}

\section{Populating the HR diagram with CSs of Galactic PNe.}

The Gaia DR2 allowed, for the first time, to determine the CS distances to a considerable sample of Galactic PNe. In order to study these CSs on an HR diagram, we selected published Johnson V magnitudes and effective temperatures of the CSs from the literature, matching the search with PNe with known $p$. For this initial study, we limited the CSs to those with V magnitudes taken from Tylenda et al. (1991, hereafter T91). We also selected only He~II Zanstra temperatures from a group of references that calculated them with the same method (Kaler 1983, hereafter K83; Shaw and Kaler 1989, hereafter SK89; and Stanghellini et al. 2002, hereafter S02). If the magnitude was not available in T91 we used the ones given by K83 or SK89. 

In Fig.~7 we show the CSs of Galactic PNe on the  ${\rm log}(T_{\rm eff}) - {\rm log}(L/L_{\odot})$ plane. In order to gauge their evolutionary stage and their mass, we superimpose the evolutionary tracks are from Vassiliadis \& Wood (1994)'s hydrogen-burning post-AGB tracks for solar metallicity ($Z=0.016$) and CS masses of 0.57, 0.6, 0.63, 0.68, 0.75, and 0.9 $M_{\odot}$, corresponding to turnoff masses of 1.0, 1.5, 2.0, 2.5, 3.5, 5.0 $M_{\odot}$. 

In Fig.~7 we also indicate the locus of the HR track by Miller Bertolami (2016), with initial mass of 1.0 $M_{\odot}$, the same of the least massive initial mass in the Vassiliadis \& Wood (1994) sample.  The slight difference in the 1.0 $M_{\odot}$ tracks is due to different CS (final) mass, which is 0.53 $M_{\odot}$ in the Miller Bertolami case, as opposed to 0.57 $M_{\odot}$ in Vassiliadis and Wood's track. This difference is due to differences in the mass-loss treatment. Since identical post-AGB mass tracks do not exist, a perfect comparison of different track sets is not possible, but we do not expect that this would change any of our conclusions of this section.

We only plot those stars whose parallax and effective temperature uncertainties are smaller than 20$\%$. The input effective temperatures (Col~3), magnitudes and their quality, if available, as in T91 (Col~4), luminosities (Col~5), and derived CS mass (Col~7) of the plotted stars are listed in Table~4. In Column(8) of the same table we give the reference code, with the first code referring to the effective temperature and the second code to the magnitude.  

The sample of 39 CSs represented in Fig.~7 is not a complete sample, and it is skewed toward nearby PNe, both for the choice of using Gaia parallaxes and for the selection of available magnitudes.  This study  is an assessment of the realm of CSs and their evolution, to confirm that Gaia distances produce a reasonable distribution of stars in the evolutionary stage that we think PN nuclei should be at, as compared to the post-AGB evolutionary tracks. Without making any further selection or other assumptions, we found that most of these CSs are correctly encompassed by the classical evolutionary tracks for H-burning post-AGB stars.

It is possible that after the final Gaia release and a careful analysis of their temperatures and luminosities there will be enough parallaxes to perform a similar study based on a CS sample that is complete within a representative distance (e.g., all PNe within 5 kpc), or for a luminosity-limited sample.  At this time, this is just an assessment of the quality of the distances, and a confirmation of the actual post-AGB nature of the CSs.

\begin{figure}
   \centering
  \includegraphics[width=12truecm]{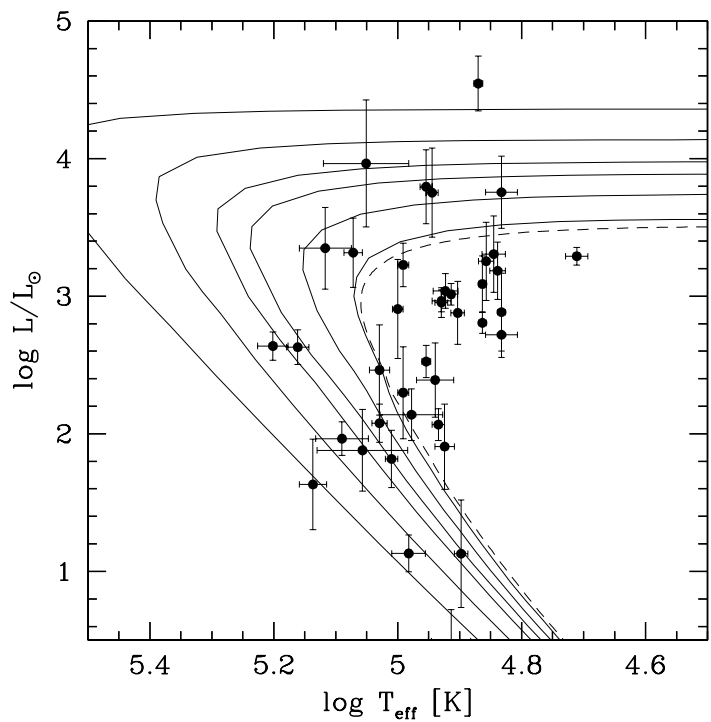}
\caption{The CSs located on the HR diagram, derived from their published magnitudes, effective temperatures, and Gaia parallaxes. Solid lines: Vassiliadis \& Wood (1994)'s evolutionary tracks for initial (i.e., turnoff) masses of 1.0, 1.5, 2.0, 2.5, 3.5, 5.0 $M_{\odot}$ and solar metallicity.  The broken line track is the 1.0 $M_{\odot}$ (turnoff mass) track by Miller Bertolami (2016).}
\end{figure}

\section{Discussion and outlook}

We presented a parallax catalog of spectroscopically-confirmed, Galactic PNe (Table~1), where we matched their CSs with the DR2 Gaia catalog. This catalog also includes published and newly measured parameters for the PNe with Gaia parallax counterparts, including morphology, angular diameters, $H_{\beta}$ fluxes, extinction constants, and linear radii.

We compared the parallaxes to all reliable individual PN parallax and distances available in the literature, to determine the accuracy of other methods. We found a good agreement between the two sets of parallaxes, including those derived by inverting other accurate distances. 

We then used the Gaia parallaxes to calibrate the most commonly used distance scales and examine in which conditions these calibrations would hold. The statistical method on which our calibration is based does use the Gaia parallaxes and their uncertainties directly, with their assumed gaussian distribution, naturally incorporating negative parallaxes. 

We determine a distance scale (Eq.~1), based on the correlation between the nebular physical radius and its surface brightness, that yield to very accurate distances for PNe whose ionized mass is not extremely low. The accuracy in terms of average distance ratio K, which is the  average of distances from the scale multiplied by Gaia DR2 parallaxes, with K=1 for scales that reproduce the DR2 parallaxes exactly, and $\sigma_{\rm K}$ representing the fractional uncertainty in the distances from the studied scale, is $<K>\pm<\sigma_{\rm K}>=0.9478\pm0.2483$ for the $|\sigma_{\rm p}/p|<0.2$ sample. 

The scatter of {\it Round} PNe on the ${\rm log} (R_{\rm pc})$ vs. ${\rm log}(Sb_{\rm H\beta})$ plane is smaller than for the general sample. This is important in assessing the distance scale, since it  proves that the basic relation is sound from a physical point of view, for PNe that are not compact, faint, or with high extinction and low ionized mass. The scale calibrated on {\it Round} PNe exclusively is excellent, but does not improve on the Eq.~(1) scale.

We also found that the ionized mass vs. optical thickness distance scale is not well constrained by Gaia parallaxes; we determined that {\it Round} PNe in the optically thin sequence of this scale define the sequence fairly well. The scale has provided guidance for parameter limits of the surface brightness- physical radius scale, but ultimately does not work as well as a distance scale for Galactic PNe.

As a working conclusion, we found that PN distances whose parallaxes are not available from Gaia or from other parallax measurements, should preferably be derived with the physical radius vs. ($H\beta$) surface brightness statistical distance scale of Eq.~(1). If the PN is not spherically symmetric, an effort should be made to derive the PN radius photometrically on the emission-line image. The scale of Eq.~(1) represents a considerable improvement over previously used statistical distance scales for Galactic PNe.

We located the CSs of the PNe on the HR diagram, by using Gaia parallaxes, and effective temperatures and magnitudes from the literature. By comparing their location to post-AGB evolutionary tracks, we determined their masses, listed in Table~4. The average mass recovered from this non-complete and non-homogeneous sample is $\sim0.62~M_{\odot}$. PNe and their CSs with well-defined parallaxes open up a broad realm of scientific possibilities. Analysis of sizable CS samples that are complete within a given luminosity could be accessible in the near future, certainly with the final Gaia release.

\acknowledgements{
We thank an anonymous Referee for their suggestions that improved this paper.
We wish to thank Eloisa Poggio for fruitful discussions on the bayesian approach.
This work has made use of data from the European Space Agency (ESA) mission
{\it Gaia} (\url{https://www.cosmos.esa.int/gaia}), processed by the {\it Gaia}
Data Processing and Analysis Consortium (DPAC,
\url{https://www.cosmos.esa.int/web/gaia/dpac/consortium}). Funding for the DPAC
has been provided by national institutions, in particular the institutions
participating in the {\it Gaia} Multilateral Agreement.
Finally B.~B., M.~G.~K., and R.~M. acknowledge the support of
the ASI contract to INAF (PI Lattanzi) n. 2018-24-HH.0.
}

 \begin{deluxetable}{ l l r r r r r r r}    
 \tabletypesize{\scriptsize}
\rotate
\tablecolumns{10}
\tablewidth{0pc}
\tablecaption{Catalog of  DR2 parallaxes and ancillary parameters for Galactic PNe\tablenotemark{a}}            
\tablehead {
\colhead{PN G}& \colhead{Gaia ID}& \colhead{$p$}&  \colhead{$M$}& \colhead{$\theta$}& \colhead{${\rm log}(F_{\rm H\beta})$}& \colhead{$c_{\alpha}$} &\colhead{${\rm log}(R_{\rm pc})$}&  \colhead{Ref.} \\
\colhead{}& \colhead{}& \colhead{[mas]}&  \colhead{$\tablenotemark{b}$}& \colhead{[$\arcsec$]}& \colhead{[erg cm$^{-2}$ s$^{-1}$]}& \colhead{}& \colhead{[pc]}& \colhead{\tablenotemark{c}}\\
\colhead{(1)}& \colhead{(2)}& \colhead{(3)}& \colhead{(4)}& \colhead{(5)}& \colhead{(6)}& \colhead{(7)}& \colhead{(8)}& \colhead{(9)}\\

}
  \startdata 
  
      000.1+02.6& 4061303281130808448& 0.5651$\pm$0.2195& $\dots$&  4.50&$\dots$&$\dots$&$\dots$&5,*,*\\
    000.3-02.8& 4056252880610683520& 0.2412$\pm$0.0957& $\dots$&$\dots$&-14.00$\pm$ 0.30&  2.07$\pm$ 0.10&$\dots$&*,5,5\\
    000.6-01.3& 4056579534283223680& 0.1277$\pm$0.1177& $\dots$&  1.50&-14.60$\pm$ 0.30&  3.79$\pm$ 0.10& -1.245$^{+ 1.185}_{- 0.381}$&5,5,5\\
    000.6-02.3& 4056324962986093184& 0.1436$\pm$0.0710& $\dots$&$\dots$&-13.20$\pm$ 0.10&  1.86$\pm$ 0.10&$\dots$&*,5,5\\
    000.7-03.7& 4050168629923554944& 1.3459$\pm$0.3993& $\dots$&  3.20&-12.61$\pm$ 0.01&  0.96$\pm$ 0.10& -1.938$^{+ 0.232}_{- 0.210}$&4,4,4\\
    001.2-03.0& 4062301564840251520& 0.1212$\pm$0.1076& $\dots$&$\dots$&-12.61$\pm$ 0.07&  1.74$\pm$ 0.10&$\dots$&*,4,5\\
    001.3-01.2& 4063378574972871680& 0.7076$\pm$0.3348& $\dots$&  2.30&-13.96$\pm$ 0.01&  3.56$\pm$ 0.10& -1.802$^{+ 0.357}_{- 0.265}$&4,4,5\\
    001.4-03.4& 4050281914036256768& 0.8155$\pm$0.0839& $\dots$&  6.50&-13.32$\pm$ 0.06&$\dots$&$\dots$&5,5,*\\
    001.6-01.3& 4063399907990515456& 0.5619$\pm$0.1684& $\dots$&  2.25&-13.90$\pm$ 0.30&  3.37$\pm$ 0.10& -1.712$^{+ 0.234}_{- 0.211}$&5,5,5\\
    001.8-03.8& 4050366641022497920& 0.5113$\pm$0.0525& $\dots$&  6.00&-12.61$\pm$ 0.02&  0.44$\pm$ 0.10& -1.245$^{+ 0.126}_{- 0.139}$&5,5,5\\

     \enddata
     \tablenotetext{a}{This is a sample of the table that will be published electronically in its completeness.}
       \tablenotetext{b}{Morphological class codes are: 1: {\it Round}, 2: {\it Elliptical}; 3: {\it Bipolar Core}; 4: {\it Bipolar}; 5: {\it Pointsymmetric}.}
     \tablenotetext{c}{The three number sequence correspond to the references for the angular radius $\theta$, the flux at $H\beta$, and the extinction constant respectively. An asterisk (*) mark the missing reference. The numbers correspond to the following references:
     1: This study; 2: Stanghellini et al. (2016); 3: Tylenda  et al. (2003); 4: CKS; 5: Acker et al. (1992). }
\end{deluxetable}

\begin{deluxetable}{ l l l l }    
\tabletypesize{\scriptsize}

\tablecaption{Comparison of DR2 parallaxes with other independent measurements}             
\tablehead{
\colhead{PN~G}& \colhead{$p$} & \colhead{$p_{\rm i}$}&  \colhead{Ref.}\\
\colhead{}& \colhead{[mas]}&  \colhead{[mas]}&  \colhead{}\\
\colhead{(1)}& \colhead{(2)}& \colhead{(3)}& \colhead{(4)}\\

}
\startdata

010.8-01.8&  0.541$\pm$0.069&  0.392$\pm$0.176&    SBJ \\
025.3+40.8&  0.380$\pm$ 0.079&  0.333$\pm$ 0.400&    SBJ\\

036.1-57.1&  4.976$\pm$ 0.076&  4.566$\pm$ 0.490&    H07\\

037.7-34.5&  0.867$\pm$ 0.116&  0.667$\pm$ 0.233&    SBJ\\

054.1-12.1&  0.407$\pm$ 0.051&  0.690$\pm$ 0.310&    SBJ\\

060.8-03.6&  2.657$\pm$ 0.044&  2.639$\pm$ 0.330&    H07\\

063.1+13.9&  1.271$\pm$ 0.059&  1.420$\pm$ 0.550&    H07\\

064.7+05.0&  0.586$\pm$ 0.063&  0.373$\pm$ 0.302&  HTB93\\

072.7-17.1&  1.474$\pm$ 0.156&  1.330$\pm$ 0.630&    H07\\
083.5+12.7&  0.635$\pm$ 0.047&  0.645$\pm$ 0.129&    SBJ\\

089.3-02.2&  0.386$\pm$ 0.027&  0.400$\pm$ 0.400&   HW88\\
096.4+29.9&  0.615$\pm$ 0.071&  0.538$\pm$ 0.081&    SBJ\\

111.0+11.6&  3.115$\pm$ 0.050&  3.333$\pm$ 0.560&    H07\\
125.9-47.0&  3.215$\pm$ 0.076&  3.356$\pm$ 0.620&    H07\\

158.9+17.8&  3.174$\pm$ 0.082&  2.740$\pm$ 0.310&    H07\\

205.1+14.2&  1.860$\pm$ 0.081&  1.848$\pm$ 0.510&    H07\\
206.4-40.5&  0.823$\pm$ 0.063&  0.433 &    C99\\
215.2-24.2&  0.645$\pm$ 0.054&  0.870$\pm$ 0.174&    SBJ\\
215.5-30.8&  2.020$\pm$ 0.072&  1.479$\pm$ 0.410&    H07\\
215.6+03.6&  0.686$\pm$ 0.030&  0.943$\pm$ 0.321&    G86\\
217.1+14.7&  1.427$\pm$ 0.129&  1.919$\pm$ 0.340&    H07\\
261.0+32.0&  0.682$\pm$ 0.088&  0.870$\pm$ 0.130&    SBJ\\
272.1+12.3&  1.157$\pm$ 0.050&  1.299 &    C99\\
285.7-14.9&  0.288$\pm$ 0.045&  0.500$\pm$ 0.150&    SBJ\\
327.8+10.0&  0.507$\pm$ 0.067&  0.588$\pm$ 0.176&    SBJ\\

\enddata
\tablecomments{SBJ: Expansion distances, Sc\"onberner et al.~(2018); H07: Trigonometric parallaxes, Harris et al.~(2007); HTB93: Expansion distances, Hajian et al.~(1993); HW88: Extinction distances, Huemer et al.~(1988); C99: Spectroscopic binaries, Ciardullo et al.~(1999); G86: Extinction distances, Gathier et al.~(1986).}
\end{deluxetable}

\begin{deluxetable}{  l l r r  r r}    
 \tabletypesize{\scriptsize}
\tablecaption{Analysis of our distance scales}             

\tablehead{ \colhead{Scale}&  \colhead{Sample}& \colhead{N} & \colhead{$\alpha_{\rm s}$} &\colhead{$<K>$}&
\colhead{$<\sigma_{\rm K}>$}\\
\colhead{(1)}& \colhead{(2)}& \colhead{(3)}& \colhead{(4)}& \colhead{(5)}& \colhead{(6)}\\

}

\startdata
Eq.~(1)&	All&			         243     & 0.1764       &  0.9823   &   0.4315	\\		
Eq.~(1)&   $|\sigma_{\rm p}/p|<0.2$&		 94   &    0.3513     &    0.9478    &   0.2483 \\
Eq.~(1)& {\it Round}&                       33&  0.2478&  1.053&  0.4289\\
Eq.~(1)& {\it Round}, $|\sigma_{\rm p}/p|<0.2$&  16& 0.1883& 0.9431& 0.2517\\

Eq.~(2)&	All&				243      &0.2184      &  1.072   &   0.4687		\\
Eq.~(2)&  $|\sigma_{\rm p}/p|<0.2$&		94    &   0.3802       &  1.06     &  0.28\\
Eq.~(2)&	{\it Round}&			      33    &  0.2616  &   1.134  &    0.4605 \\
Eq.~(2)& {\it Round}, $|\sigma_{\rm p}/p|<0.2$&   16   &   0.1358  &      1.032  &     0.2775\\

 \\
 
\enddata
\end{deluxetable}

\begin{deluxetable}{  l l r r l r l r}    
\tabletypesize{\scriptsize}

\tablecaption{Temperature, luminosity, and mass of CSs}             

\tablehead{ \colhead{PN~G}&  \colhead{Name}& \colhead{${\rm log}(T_{\rm eff})$} & \colhead{V} & \colhead{Q}& \colhead{${\rm log}(L/L_{\odot})$}&
\colhead{Ref}& \colhead{$M_{\rm CS}$}\\
\colhead{(1)}& \colhead{(2)}& \colhead{(3)}& \colhead{(4)}& \colhead{(5)}& \colhead{(6)}& \colhead{(7)}& \colhead{(8)}\\
}
\startdata
   017.3-21.9    &      A~65   &     4.940$\pm$0.030&        15.9 &	 NA   &    2.391$\pm$0.270&		  K83, K83&	$<$0.57\\
     036.0+17.6   &       A~43   &    4.833$\pm$0.006  &     14.75  &     A &      2.884$\pm$0.284&		K83, T91& $<$0.57\\
     036.1-57.1 &     NGC~7293  &       5.090$\pm$0.043  &     13.5    &    A   &    1.964$\pm$ 0.122&		K83, T91&	0.715\\
     041.8-02.9  &     NGC~6781  &      4.982$\pm$0.027  &     16.78    &    B  &      1.130$\pm$0.134	&	K83, T91&	0.825\\
     045.7-04.5 &      NGC~6804 &       4.954$\pm$0.007    &  14.37    &	 A  &     2.525 $\pm$0.118	&	S02, T91&	$<$0.57\\
    047.0+42.4  &         A~39       & 4.934$\pm$0.010      &  15.69    &   A    &   2.066$\pm$0.116&		K83, T91&	$<$0.57\\
     059.7-18.7  &         A~72       & 5.029$\pm$0.016      & 16.12	&NA &      2.463$\pm$0.327&		K83, T91&      0.57\\ 
     060.8-03.6 &      NGC~6853  &      5.201$\pm$0.025   &    13.94  &      B&       2.638$\pm$0.103			&K83, T91&	0.715\\
     061.4-09.5  &     NGC~6905   &     5.117$\pm$0.042       &15.7       &	 C &      3.349$\pm$0.297	&	S02, T91&	0.585\\
     063.1+13.9 &      NGC~6720   &     5.161$\pm$0.018    &   15.29    &    B  &     2.629$\pm$0.127	&	K83, T91& 	0.655\\
     066.7-28.2   &    NGC~7094   &     4.863$\pm$0.006     &  13.68   &    B  &     3.088$\pm$0.204	&	K83, T91&		0.57\\
     081.2-14.9    &       A~78       & 4.839$\pm$0.013       & 13.21   &    A    &   3.185$\pm$0.207 	&	K83, T91&	$<$0.57\\
     084.9+04.4&           A~71      &  5.137$\pm$0.022      & 18.95     &   NA   &    1.632$\pm$0.330	&	K83, T91&	0.9\\
     094.0+27.4  &       K~1-16    &    4.903$\pm$0.011      & 15.08   &     A    &   2.878$\pm$0.229	&	K83, T91&	$<$0.57\\   
     106.5-17.6  &     NGC~7662  &      5.051$\pm$0.069    &   13.2 &      	 D    &   3.964$\pm$0.462	&	K83, T91& 0.68\\
     118.8-74.7  &     NGC~246    &    4.929$\pm$0.010   &    11.96  &     	 A  &     2.954$\pm$0.109	&	K83, T91& $<$0.57\\
     144.5+06.5 &      NGC~1501 &       4.923$\pm$0.020    &  14.39  &     	B  &     3.038$\pm$0.126	&	S02, T91&	$<$0.57\\
     148.4+57.0    &   NGC~3587  &      5.029$\pm$0.012  &     16.01&      	 B   &    2.077$\pm$0.140	&	K83, T91& 0.615\\
     158.8+37.1 &          A~28     &   4.914$\pm$0.032  &     17.4  &     	 NA   &   0.493$\pm$0.230&		K83, T91& $>$0.9\\ 
     164.8+31.1  &       Jn~Er~1   &      5.010$\pm$0.010    & 16.83      	 &A      & 1.817$\pm$0.209	&	S02, T91&	0.63\\
     165.5-15.2   &    NGC~1514   &     4.711$\pm$0.018  &     9.42       &	A   &    3.291$\pm$0.064	&		S02, T91& $<$0.57\\
     189.1+19.8   &    NGC~2372  &      5.072$\pm$0.015 &      14.85   &  	A     &  3.317$\pm$0.251	&	K83, T91& 0.57\\
     197.8+17.3   &    NGC~2392   &      4.870$\pm$0.007   &   10.53       & 	A      & 4.546$\pm$0.199	&	S02, T91&  $>$0.9\\
     208.5+33.2   &        A~30   &     4.857$\pm$0.011   & 14.38        & 	 A      & 3.254$\pm$0.240	&	K83, T91& $<$0.57\\
 
     214.9+07.8   &        A~20    &    4.991$\pm$0.009 &      16.56     & 	NA   &    2.299$\pm$0.334&		K83, K83& $<$0.57\\
     219.1+31.2  &         A~31    &    5.057$\pm$0.073     &  15.51      & NA       & 1.880$\pm$0.297	&	K83, K83&0.68\\
     220.3-53.9   &    NGC~1360    &    4.929$\pm$0.015   &    11.35      &  NA      & 2.965$\pm$0.080	&	K83, K83& $<$0.57\\
     238.0+34.8 &          A~33    &    4.978$\pm$0.051     & 15.5       &	 A     &  2.139$\pm$0.188	&	SK89, T91&0.57\\
     239.6+13.9  &     NGC~2610    &        5.000$\pm$0.009    &    15.9     &	A       &2.907$\pm$0.360	&	SK89, T91&$<$0.57\\
 
     261.0+32.0   &    NGC~3242   &     4.954$\pm$0.010   &    12.31   &    B   &    3.795$\pm$0.270	&	SK89, T91&0.615\\
     285.7-14.9   &     IC~2448     &   4.944$\pm$0.010      & 14.22   	&B      & 3.752$\pm$0.324	&	SK89, T91&0.615\\
     288.8-05.2 &      He~2-~51    &    4.833$\pm$0.026      & 15.66     &   A  &     2.719$\pm$0.164	&	SK89, T91&$<$0.57\\
     294.1+43.6   &    NGC~4361    &    4.991$\pm$0.009   &    13.21     &  A   &    3.227$\pm$0.158  	&	SK89, T91&	0.57\\
   300.5-01.1 &    He~2-85&       4.898$\pm$0.011&    16.59 & B&   1.128$\pm$0.392 &  SK89, T91&  0.63\\
   
     318.4+41.4   &        A~36     &   4.863$\pm$0.006      & 11.53       & A     &  2.806$\pm$0.077	&	SK89, T91&	$<$0.57\\
     327.8+10.0  &     NGC~5882    &    4.845$\pm$0.019     &  13.43   &      B   &    3.306$\pm$0.277 	&	SK89, T91&	$<$0.57\\
     329.0+01.9   &        Sp~1   &     4.914$\pm$0.011     &14.03        & A    &   3.014$\pm$0.078	&	SK89, T91&	$<$0.57\\
     341.6+13.7 &      NGC~6026     &   4.833$\pm$0.026    &   13.29   &      A   &    3.755$\pm$0.263	&	SK89, T91&	0.6\\
 
 \enddata
 \tablecomments{reference codes are: K83: Kaler 1983; T91: Tylenda et al. 1991; S02: Stanghellini et al. 2002; SK89: Shaw \& Kaler 1989.}

 \end{deluxetable}

\appendix
\section{Maximum likelihood method for the physical nebular radius vs. $H\beta$ surface brightness distance scale}

The statistical distance scale that we describe in $\S$3.1 relates linearly the logarithm of the physical radius to that of the $H\beta$ surface brightness of the PNe, i.e.:
$${\rm log}(R_{\rm pc})=a \times{\rm log}(Sb_{\rm H\beta}) + b, \eqno(A1).$$

For the i-th calibrator target, we have a set of measurements ($p_{\rm i}$, $\theta_{\rm i}$, $I_{\rm i}=F_{\rm H\beta, i}\times10^{c_{\rm i}}$) of the variables
$\omega_{\rm i}$ (parallax), $\phi_{\rm i}$ (angular radius), and 
$J_{\rm i}$ (extinction-corrected flux, or intensity, at $H\beta$),
where $R=\phi/(206265\,\omega)$ and $Sb_{\rm H\beta}=J/\phi^2$, with $\phi$ and $\omega$ in arcsec and $J$ in flux units of erg cm$^{-2}$s$^{-1}$.

By eliminating logarithms, and solving for the parallax variable, we obtain the following nonlinear relation:
$$\omega=\frac{\pi^a\phi^{2a+1}}{206265\, J^a 10^b}+\epsilon, \eqno(A2)$$
where $\epsilon$ represents the intrinsic scatter of the relation. Our goal is to estimate {\it a} and {\it b}, given the measurements $p_{\rm i}$, $\theta_{\rm i}$, $I_{\rm i}$ and their uncertainties. We approach the problem by marginalizing the complete likelihood function of the i-th set of measurements with respect to the variables $\omega$, $\phi$, $J$, which are also characterized by an a-priori probability function (e.g., Kelly 2007 and references therein). We choose $\omega$ as dependent variable and express its conditioned probability function as $p(\omega|\phi,J)$. Thus we can write the marginalized likelihood as:

$$p(p_{\rm i},\theta_{\rm i},I_{\rm i}|a,b)=\int\int\int_{0}^{\infty} p(p_{\rm i},\theta_{\rm i},I_{\rm i},\omega_{\rm i},\phi_{\rm i},J_{\rm i}|a,b) \,d\omega \,d\phi\, dJ=\]
\[\int\int\int_{0}^{\infty}p(p_{\rm i},\theta_{\rm i},I_{\rm i},|\omega_{\rm i},\phi_{\rm i},J_{\rm i}, a, b)p(\omega_{\rm i}|\phi_{\rm i},J_{\rm i})p(\phi_{\rm i})p(J_{\rm i})\,d\omega \,d\phi \,dJ\ .\eqno(A3)$$

All measurements are gaussian-distributed random variables, i.e., $\theta_{\rm i}\sim N(\phi_{\rm i},\sigma_{\theta_{\rm i}})$,
$I_{\rm i}\sim N(J_{\rm i},\sigma_{\rm I_i})$, and 
$p_{\rm i}\sim N(\omega_{\rm i}, \sigma_{p_{\rm i}})$; since  they are also uncorrelated, the likelihood is split in the product of single probabilities:

$$p(p_{\rm i},\theta_{\rm i},I_{\rm i} | a,b)= \int\int\int_{0}^{\infty}  N(\frac{\pi^a\theta_i^{2a+1}}{206265\, J_i^a  10^b},\sigma_{\omega_i})N(\theta_{\rm i},\sigma_{\theta_i})N(J_{\rm i},\sigma_{I_{\rm i}})p(\omega_i)p(\theta_i)p(J_{\rm i})\,d\omega \,d\theta \,dJ .\eqno(A4)$$

In general we can express $N(x,\sigma_x)=\frac{1}{\sqrt{2\,\pi}\sigma_{x_{m}}} exp (-\frac{(x_{m}-x)^2}{2\,\sigma^2_{x_{m}}})$, where we indicate with the suffix {\it m} the measured value. If we disregard the intrinsic scatter of the distance scale, we can express the conditional parallax probability as  $p(\omega|\phi,J)=\delta(\omega-\frac{\pi^a\phi^{2a+1}}{206265\, J^a 10^b})$, which means that for each pair of measured $(\phi, J)$,  $\omega \neq 0$ only in a singularity. Therefore, the integration in $\omega$  disappears, transforming the likelihood function into:

$$p(p_{\rm i},\theta_{\rm i},I_{\rm i}|a,b)=\int\int_{0}^{\infty} N(\frac{\pi^a\phi_{\rm i}^{2a+1}}{206265\, J_{\rm i}^a  10^b},\sigma_{p_{\rm i}})N(\phi_{\rm i},\sigma_{\theta_{\rm i}})N(J_{\rm i},\sigma_{I_{\rm i}})p(\phi_{\rm i})p(J_{\rm i})\,d\phi \,dJ.\eqno(A5)$$

The complete likelihood for the set of calibrators is the product of the likelihoods of each target. If we assume that the probability densities of {\it a} and {\it b}, i.e., $p(a)$ and $p(b)$, are uniformly distributed, the posterior distribution of {\it a} and {\it b} for the realization of the variables $\omega$, $\phi$, and $J$ is proportional to the likelihood function of the data (see Bayes theorem), i.e.:

$$p(a,b|p_{\rm i},\theta_{\rm i},I_{\rm i}) \propto \prod_{i=1}^N p(p_{\rm i},\theta_{\rm i},I_{\rm i}|a,b), \eqno(A6)$$

where $p(p_{\rm i},\theta_{\rm i},I_{\rm i}|a,b)$ is the evidence for the i-th measurement, and N is the total number of targets. If we express Eq.~(A6) in logarithmic form, and use the explicit format of the right hand side, we obtain that
the probability density for the realization of $a$ and $b$, conditioned by the measurements $(p_{\rm i},\theta_{\rm i},I_{\rm i})$, is:

$$p(a,b|p_{\rm i},\theta_{\rm i},I_{\rm i},{\bf P}) \propto \sum_{i=1}^N log[\int\int_{0}^{\infty} N(\frac{\pi^a\phi_{\rm i}^{2a+1}}{206265\, J_{\rm i}^a 10^b},\sigma_{p_{\rm i}})N(\phi_{\rm i},\sigma_{\theta_{\rm i}})N(J_{\rm i},\sigma_{I_{\rm i}})p(\phi_{\rm i})p(J_{\rm i})\,d\phi \,dJ]. \eqno(A7)$$

We choose the integration extremes $\phi_{\rm lim}$ and $J_{\rm lim}$ to be compatible with the observations, i.e., $\pm 5\sigma$ for each variable, and the probability densities $p(\phi)$ and  $p(J)$ to be uniform between these chosen limits. It follows that:

$$p(a,b|p_{\rm i},\theta_{\rm i},I_{\rm i}) \propto \sum_{k=1}^N log[\int_{0}^{\phi_{lim}} \int_{0}^{J_{lim}}N(\frac{\pi^a\phi_{\rm i}^{2a+1}}{206265\, J_{\rm i}^a 10^b},\sigma_{p_{\rm i}})N(\phi_{\rm i},\sigma_{\theta_{\rm i}})N(J_{\rm i},\sigma_{I_{\rm i}})\,d\phi \,dJ].\eqno(A8)$$

We performed a numerical solution to this problem, for a grid of $a$ and $b$ values to build the bi-variate
{\it posterior probability function}, $p(a,b)$, from which we determined the maximum likelihood estimates of $a$ and $b$ and their confidence intervals. As an example, Fig.~8 shows the computed likelihood plot for the case of $|\sigma_{\rm p}/p|<0.2$.

\begin{figure}
   \centering
  \includegraphics[width=\hsize]{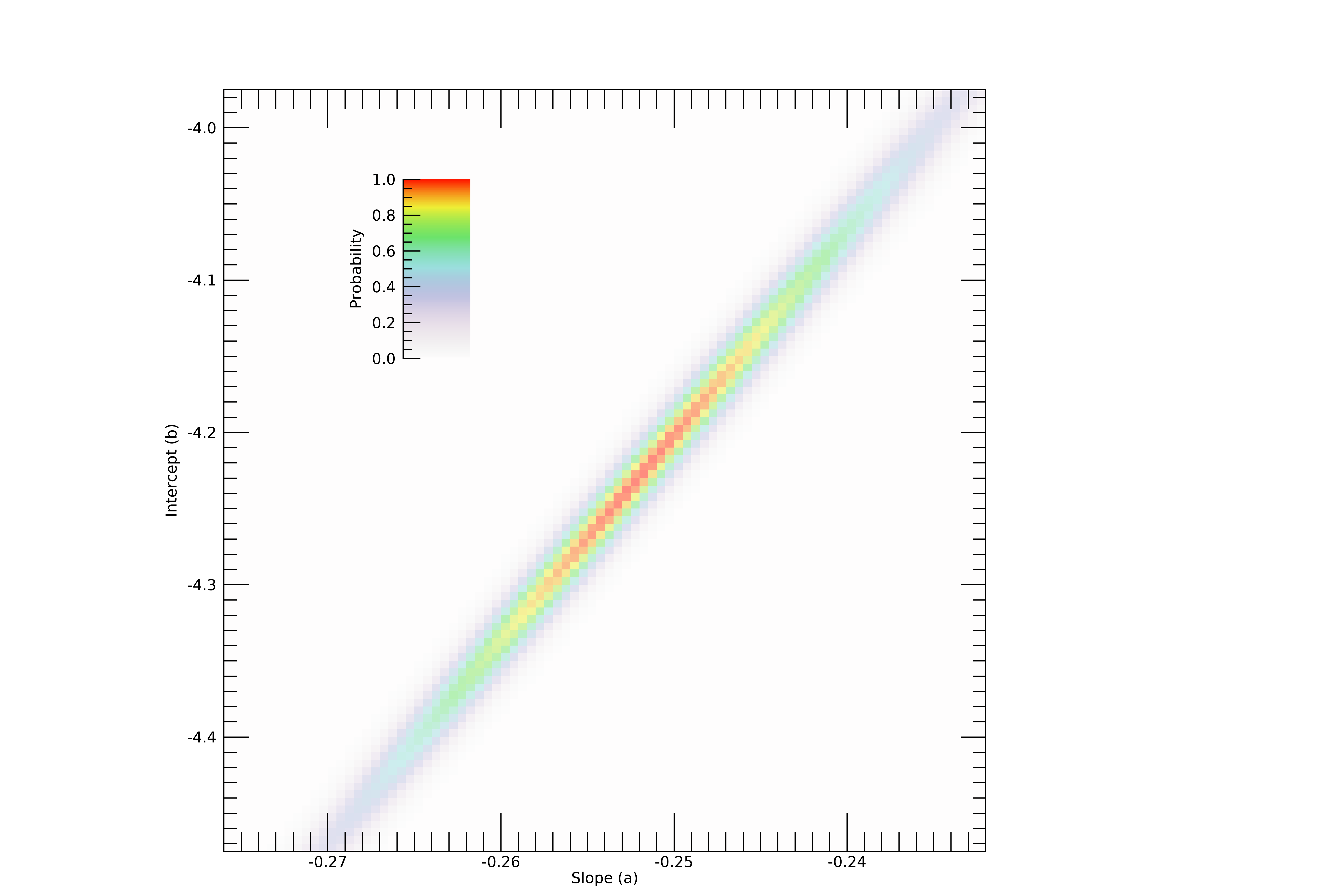}
\caption{Plot of the posterior probability distribution for the {\it a}
(slope) and {\it b} (intercept) distance scale parameters, case of PNe with
$|\sigma_{\rm p}/p|<0.2$.}
\end{figure}

\end{document}